\documentclass{article}

\usepackage{arxiv}

\usepackage[utf8]{inputenc} % allow utf-8 input
\usepackage[T1]{fontenc}    % use 8-bit T1 fonts
\usepackage{hyperref}       % hyperlinks
\usepackage{url}            % simple URL typesetting
\usepackage{amsmath}
\usepackage{multirow}
\usepackage{listings}
\usepackage{booktabs}       % professional-quality tables
\usepackage{amsfonts}       % blackboard math symbols
\usepackage{nicefrac}       % compact symbols for 1/2, etc.
\usepackage{microtype}      % microtypography
\usepackage{cleveref}       % smart cross-referencing
\usepackage{lipsum}         % Can be removed after putting your text content
\usepackage{graphicx}
\usepackage{natbib}
\usepackage{doi}
\usepackage{subfigure}
\usepackage{caption}
\usepackage{authblk}
\usepackage{tcolorbox} % For boxed code

\newtcolorbox[auto counter, number within=section]{listing}[2][]{colframe=black!10!white, colback=black!5!white, coltitle=black, fonttitle=\bfseries, #1} 
\title{PainDECOG: Machine Learning-Based Identification of Pain Biomarkers from sEEG Signals}

\author[1]{Sidharth\textsuperscript{\textdagger}}
\author[2]{Vishwas Sathish\textsuperscript{\textdagger}}
\author[2]{Shweta Bansal}
\author[3]{Samantha Sun}
\author[3]{Timmy Pham}
\author[4]{Kurt Weaver}
\author[1,2,3]{Rajesh P. N. Rao}
\author[1,2,3,5]{Jeffrey Herron}

\affil[1]{Department of Electrical and Computer Engineering, University of Washington, Seattle}
\affil[2]{Paul G. Allen School of Computer Science and Engineering, University of Washington, Seattle}
\affil[3]{Department of Bioengineering, University of Washington, Seattle}
\affil[4]{Department of Radiology, University of Washington, Seattle}
\affil[5]{Department of Neurological Surgery, University of Washington, Seattle}

\begin{document}

\maketitle
\def\thefootnote{\textdagger}\footnotetext{These authors contributed equally to this work. }
% Correspondence should be addressed to Sidharth (\textit{sid17@uw.edu})}\def\thefootnote{\arabic{footnote}}
% \abstract*{Each chapter should be preceded by an abstract (no more than 200 words) that summarizes the content. The abstract will appear \textit{online} at \url{www.SpringerLink.com} and be available with unrestricted access. This allows unregistered users to read the abstract as a teaser for the complete chapter.
% Please use the 'starred' version of the \texttt{abstract} command for typesetting the text of the online abstracts (cf. source file of this chapter template \texttt{abstract}) and include them with the source files of your manuscript. Use the plain \texttt{abstract} command if the abstract is also to appear in the printed version of the book.}
% \vspace{-4em}
\begin{abstract}
    This study presents a systematic machine-learning approach for classifying acute pain from raw electrophysiological signals. We address binary and ternary classification tasks, leveraging Power-In-Band (PIB) and signal coherence—a functional connectivity metric—as distinguishing features. 
Our method evaluates the effectiveness of traditional machine learning algorithms on a manually curated electrophysiological dataset obtained from intracranial electroencephalography (iEEG), offering valuable insights into model performance for pain detection. Furthermore, we identify critical electrode pairings associated with acute pain, providing a clearer understanding of the neural markers that differentiate pain states. This work highlights the potential of targeted feature engineering in advancing pain classification, setting the stage for future enhancements in real-time and personalized pain assessment tools. Additionally, these findings have promising applications in neuromodulation and Deep Brain Stimulation (DBS), where adaptive and closed-loop systems could leverage identified pain markers to modulate pain-related brain regions more precisely, offering improved therapeutic options for chronic pain management
\end{abstract}

\section{Introduction}
Pain is a complex sensory phenomenon frequently encountered after surgery, with around 75\% of patients reporting moderate to severe acute pain in the postoperative period \cite{Horn2024Nov}. Due to pain's highly subjective nature and its variability based on factors such as type of stimulus, mood, attention, and memory, a universal treatment remains elusive \cite{Lumley2011Jun}. Acute pain is often managed through medications like opioids, which are potent but come with a heightened risk of misuse and addiction \cite{Vowles2015Apr}. Neuromodulation, like deep brain stimulation, has shown promise for cases where pain sources are known and accessible; however, these techniques are typically invasive and expensive \cite{Bittar2005Jun, Fana2020Feb, Shirvalkar2018Mar}. Effective pain management continues to face significant challenges due to limited mechanistic insights and the lack of reliable tools for assessing the subjective experience of pain.

Traditional models of pain delineate pathways leading to distinct brain circuits \cite{Garland2012Jul, Scholz2014Dec}. Broadly, pain follows either nociceptive or neuropathic pathways \cite{Clauw2019Apr}. In the nociceptive pathway, harmful stimuli lead to the peripheral sensitization of nociceptors, increasing action potentials and transmitting signals to the spinal cord's dorsal horn, and then to the thalamus, where central processing of pain occurs. By contrast, neuropathic pain results from direct damage to the nerves in the peripheral or central nervous system, similarly triggering central sensitization due to independent, stimulus-free activity and biomedical changes. The thalamus then routes pain signals to various regions in the cerebral cortex, producing responses influenced by emotional and environmental contexts \cite{PrinciplesofNeuralScience}.

Neuroimaging studies have identified a network of interconnected brain regions involved in pain processing, commonly referred to as the "pain matrix" \cite{Morton2016Sep}. This network includes areas such as the primary and secondary somatosensory cortices, anterior cingulate cortex, prefrontal cortex, insular cortex, amygdala, thamalus, cerebellum, and periaqueductal gray. Within this framework, it is hypothesized that the lateral pathway of the pain matrix primarily processess sensory aspects, while the medial pathway is associated with emotional aspects of pain \cite{Kulkarni2005Jun}.

% Electrophysiology studies have provided additional insights into pain processing circuitry. For instance, recent advances in electrophysiological methods, including electroencephalography (EEG), electrocorticography (ECoG), and magnetoencephalography (MEG), offer valuable perspectives on the temporal dynamics of pain. These methods enable researchers to track fluctuations in electric potential across brain regions in real time, which may represent synaptic activity and the resulting transmission of signals between brain areas \cite{Chen2021Oct, Buzsaki2012Jun}.

% Functional connectivity analysis, which examines patterns of neural activity across regions, has proven useful in neuroimaging and electrophysiology studies to explore associations with neurological conditions, cognitive functions, and behaviours \cite{Bastos2016Jan}. For example, connectivity patterns derived from this analysis have effectively distinguished individuals with Alzheimer's disease from those experiencing healthy aging, and depressed patients from non-depressed controls \cite{Greicius2004Mar, Anand2005May}. Functional connectivity analysis has also been applied to the study of epilepsy, memory processing, emotion detection and multiple sclerosis \cite{Tracy2015Apr, Stark2021Aug, sid, Tran2023Jan}. In pain research, this technique has been utilized to distinguish between different levels of heat-induced pain intensity in experimental settings, provided a nuanced understanding of pain perception \cite{Wager2013Apr}.

In this paper, we develop machine learning frameworks for classifying subject-specific binary and ternary pain states from post-surgical fluctuations in pain using signal power and functional connectivity analysis on iEEG. Our dataset includes subjects with drug-resistant epilepsy undergoing seizure monitoring, who reported varying levels of pain over their 7–10 day hospital stay. Pain intensity was recorded at unscheduled intervals using the Visual Analog Scale (VAS), a numerical measure from 0 to 10. We synchronized neural activity with these pain assessments and analyzed spectral features, such as Power-In-Band (PIB) and Magnitude Squared Coherence (MSC). We performed binary and ternary pain classification using traditional machine learning classifiers—Logistic Regression (LR; Baseline), Support Vector Machine (SVM), and Random Forest (RF). These methods helped us identify the regions of the brain for each patient that contributed most to pain classification. 
% In particular, coherence features revealed key electrode-electrode combinations instrumental in distinguishing pain levels. This approach provides insights into subject-specific pain responses, potentially informing future neuromodulation strategies.

\section{Dataset}
\label{sec:2}
% Always give a unique label
% and use \ref{<label>} for cross-references
% and \cite{<label>} for bibliographic references
% use \sectionmark{}
% to alter or adjust the section heading in the running head
% Instead of simply listing headings of different levels we recommend to let every heading be followed by at least a short passage of text.  Further on please use the \LaTeX\ automatism for all your cross-references and citations.

% Please note that the first line of text that follows a heading is not indented, whereas the first lines of all subsequent paragraphs are.

% Use the standard \verb|equation| environment to typeset your equations, e.g.
% %
% \begin{equation}
% a \times b = c\;,
% \end{equation}
% %
% however, for multiline equations we recommend to use the \verb|eqnarray| environment\footnote{In physics texts please activate the class option \texttt{vecphys} to depict your vectors in \textbf{\itshape boldface-italic} type - as is customary for a wide range of physical subjects}.
% \begin{eqnarray}
% \left|\nabla U_{\alpha}^{\mu}(y)\right| &\le&\frac1{d-\alpha}\int
% \left|\nabla\frac1{|\xi-y|^{d-\alpha}}\right|\,d\mu(\xi) =
% \int \frac1{|\xi-y|^{d-\alpha+1}} \,d\mu(\xi)  \\
% &=&(d-\alpha+1) \int\limits_{d(y)}^\infty
% \frac{\mu(B(y,r))}{r^{d-\alpha+2}}\,dr \le (d-\alpha+1)
% \int\limits_{d(y)}^\infty \frac{r^{d-\alpha}}{r^{d-\alpha+2}}\,dr
% \label{eq:01}
% \end{eqnarray}

% \section{Dataset} 
Intracranial electroencephalography (iEEG) electrodes were implanted in patients (n=4; one subject lacked a high pain distribution, so the main analysis includes three subjects, with the data and results for the fourth subject provided in the appendix) with intractable epilepsy for the purpose of seizure monitoring and identifying epileptogenic zones. All experiments were approved by the University's Institutional Review Board, and informed consent was obtained from all participants.

\begin{figure}[ht]
\centering
    \begin{tabular}{ccc}
        \subfigure[Subject 1]{
            \includegraphics[width=0.15\textwidth]{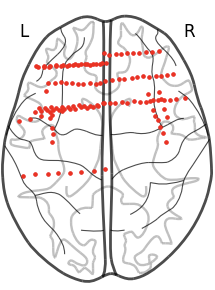}
            \label{fig:422bc5_elec}
        } &
        \subfigure[Subject 2]{
            \includegraphics[width=0.15\textwidth]{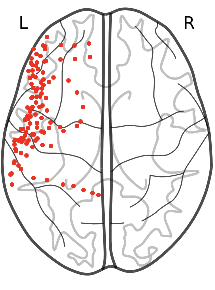}
            \label{fig:0b5a2e_elec}
        } &
        \subfigure[Subject 3]{
            \includegraphics[width=0.15\textwidth]{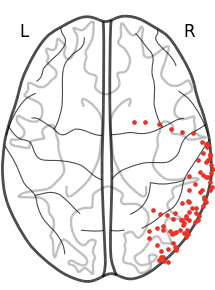}
            \label{fig:c5a5e9_elec}
        }
    \end{tabular}
    \caption{Axial view of channel coverage across subjects. Subject 2 and Subject 3 have surface ECoG electrodes concentrated in a single hemisphere, while Subject 1 has bilateral sEEG coverage.}
    \label{fig:electrode_location}
\end{figure}

Neural electrophysiological data was collected using implanted stereo-electroenc-\\-ephalography (sEEG) electrodes or subdural multi-contact electrocorticography (ECoG) signals. The sEEG electrodes (manufactured by PMT Corp, USA) feature between 8 and 16 individual channels per electrode, with a diameter of 0.8mm and a spacing of 3.5mm between channels. In contrast, ECoG grids (produced by Ad-Tech, USA) comprise 8 to 64 recording contacts, each with a diameter of 2.3mm with a spacing of 10mm. Electrode implantation sites were determined based on clinical criteria, leading to variations in electrode coverage across subjects (Figure \ref{fig:electrode_location}). 

Neural activity was recorded at 500 Hz sampling rate using up to 256 electrodes over periods ranging from 5-7 days.

The data underwent pre-processing using a notch filter to eliminate line noise at 60 Hz, as well as its harmonics at 120 Hz and 180 Hz. Given that local field potential activity typically occurs within the 0.1-200 Hz range \cite{Lashgari2012Aug}, a 5th-order Butterworth low-pass filter with a 200 Hz cutoff was applied to filter out high frequency components. The preprocessed data was then visually inspected to assess its quality to identify the recording channels that were potentially unusable due to excessive noise or artifacts.

Throughout their hospital stay, patients were asked to rate their pain intensity using Visual Analogue Scale (VAS) \cite{Langley1985Jul}. VAS contains a horizontal line that represents the varying levels of pain, with 0 indicating no pain and 10 indicating the worst pain imaginable.
Pain reports included the time of pain score reporting, intensity levels, pain descriptors (mild, moderate, severe), pain location, and patterns (none, intermittent, continuous). Pain scores were collected several times daily as part of routine clinical care, resulting in 54-68 scores per subject (refer Table \ref{tab:dataset}). 
To categorize discrete pain scores for ternary and binary classification tasks, we applied three distinct strategies, drawing on established methodology.

\textbf{Strategy 1} aligned with the thresholds defined in \cite{Boonstra2014Dec}. For ternary classification, we segmented pain scores as follows: scores $\leq 3$ indicated \textit{no pain}, scores between 4 and 6 represented \textit{moderate pain}, and scores $\geq 7$ denoted \textit{high pain}. For binary classification, we defined \textit{no pain} as any score $\leq 3$, while pain was represented by scores $> 3$.

\textbf{Strategy 2} considers the context in which the patient data was collected—post-surgery, where moderate and high pain levels are prevalent. Using a threshold of 7 to distinguish \textit{moderate pain} from \textit{high pain} allows for a more clinically meaningful classification. This approach aligns with the nature of post-operative pain assessments and enhances the ability to capture variations in pain intensity effectively.

\textbf{Strategy 3} arose from analyzing the pain score distribution across subjects (see Figure \ref{fig:pain_label_kde}
 in the supplementary section). We observed significant overlap in the distribution of \textit{moderate pain} with both \textit{low pain} and \textit{high pain}, which could potentially confuse the model during training by creating ambiguity when mapping features to labels. To address this, we hypothesized that excluding data with pain scores in the range $[4, 6]$ would reduce overlap and improve the model's ability to distinguish between the remaining classes.

\begin{table}[ht]
\centering
\begin{tabular}{|| c || c || c || c || c  || c ||} 
 \hline
 \textbf{ID} & \textbf{Total} & \textbf{Useful} & \textbf{Trials} & \textbf{Modality}  \\ 
 & \textbf{channels} & \textbf{channels} & &\\
 \hline\hline
Subject 1 &129 &128 & 54 & sEEG\\
 \hline
 % Subject 2 &129 &94 & 36& ECoG\\
 % \hline
 Subject 2 & 129  &116 & 68 & ECoG\\ 
 \hline
 Subject 3 &129 &90 &64 & ECoG \\
 % \hline
 % 6c29e3 & 200 & 198 & 16 & ECoG \\
 \hline

\end{tabular}
   \caption{Dataset description. Useful channels are obtained after discarding the channels with only zero-valued signals}
  \label{tab:dataset}
\end{table} 

Neural activity windows surrounding each pain score report were extracted and used for classifying pain states. A 5-minute window was defined, spanning 2.5 minutes before and after each pain rating, with each window considered as a separate trial. It was assumed that the subject’s pain remained relatively stable during this time. Due to the limited amount of available data, each 5-minute neural activity window was divided into thirty 10-second sub windows.

\section{Feature Extraction and Methodology}
\label{subsec:2}
To characterize pain-related neural activity, we used two main feature sets: (1) Total Power-In-Band values and (2) Magnitude Squared Coherence values.

\subsection{Power-In-Band (PIB) values}
The PIB feature set captures the power of neural signals within specific frequency bands, providing insight into the spectral characteristics associated with pain processing. Building on the approach described in \cite{Gram2015Apr, Nezam2018Jun}, which included power analysis in the delta, theta, alpha, beta, and gamma bands, we extended this feature set to also include high-gamma bands. This led to a total of six frequency bands for our analysis, covering a broad range of neural oscillatory activity potentially relevant to pain.

To compute the PIB values, we first applied a second-order Butterworth bandpass filter to isolate each target frequency band from the raw signal. For each electrode, we segmented the filtered signals into windows and calculated the power in each band by summing the squared amplitude envelope of the signal. This was achieved by applying the Hilbert transform to obtain the analytic signal, from which the power was computed as
\begin{equation}
    \text{PIB} = \sum_{t} \left| \mathcal{H}(x_{\text{filtered}}(t)) \right|^2
\end{equation}
where  $x_{\text{filtered}}(t)$  represents the bandpass-filtered signal, and $ \mathcal{H}(x_{\text{filtered}}(t))$ denotes the analytic signal produced via the Hilbert transform. This formulation provides a robust measure of total power within each frequency band for each electrode, capturing frequency-specific power changes that may correlate with pain perception

\subsection{Magnitude Squared Coherence}
Motivated by \cite{Battiti1994Jul}, which used the mutual information criterion to evaluate a set of candidate features and select an informative subset as input for a neural network classifier, we adopted a similar approach. We computed the mutual information between each electrode’s signal and pain intensity for each of the frequency bands to identify the top 20 electrodes most strongly associated with pain scores. This selection allowed us to reduce the feature set while retaining key information relevant to pain classification, optimizing computational efficiency.

Mutual information $ I(X; Y)$ between an electrode signal $X$ and pain intensity $Y$ quantifies the shared information, and is calculated as:
\begin{equation}
    I(X; Y) = \sum_{x \in X} \sum_{y \in Y} p(x, y) \log \left( \frac{p(x, y)}{p(x) p(y)} \right)
\end{equation}

where:
$p(x, y)$ is the joint probability distribution of $ X $ and $ Y $,
$p(x)$ and $p(y)$ are the marginal probability distributions of $X$ and $Y$, respectively.

After selecting the top 20 electrodes, we calculated the Magnitude Squared Coherence (MSC) values for each unique pair of electrodes within this subset.

Magnitude Squared Coherence (MSC) quantifies the linear relationship between a pair of signals across frequencies. The MSC between  $X_i$ and $X_k$ is defined as:

\begin{equation}
\text{MSC}_{i,k}(\omega) = \frac{|S_{i,k}(\omega)|^2}{S_{i,i}(\omega) S_{k,k}(\omega)}
\end{equation}

where  $S_{i,i}(\omega) $ and $ S_{k,k}(\omega) $ represent the power spectral densities of signals $X_i$ and $X_k$, respectively, and $S_{i,k}(\omega) $ is the cross-power spectral density between them. The value of MSC lie in the range $[0, 1]$.

We calculated both features across each of the six frequency bands and organized them into two separate feature spaces: a PIB feature set and a MSC feature set, each stacked independently.

% In the binary and ternary classification pipeline, we first separated 10\% of each subject's total trials as test data to mitigate the risk of data leakage between the training and test sets. The remaining trials were then subjected to a 20-fold cross-validation strategy. In each iteration, we trained a model on a randomly selected subset of trials, with each trial divided into 30 non-overlapping 10-second windows. The trained model was subsequently evaluated on the designated test data. This experimental procedure was repeated 15 times, yielding an accuracy distribution across the iterations. We report the mean accuracy for each subject, with the overall pipeline illustrated in 
The raw ECoG/sEEG signals are initially divided into 30 non-overlapping 10-second windows using a Hanning window function to maintain signal continuity. Labels are then binarized or categorized based on pain intensity levels.

\begin{figure}[ht]
    \centering
    \subfigure[]{
        \includegraphics[width=0.5\textwidth]{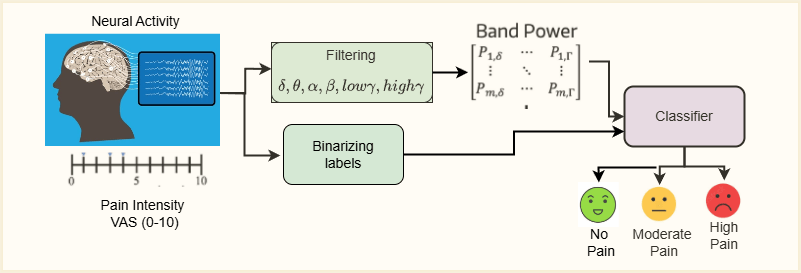}
        \label{fig:bp}
    }
    \hfill
    \subfigure[]{
        \includegraphics[width=0.5\textwidth]{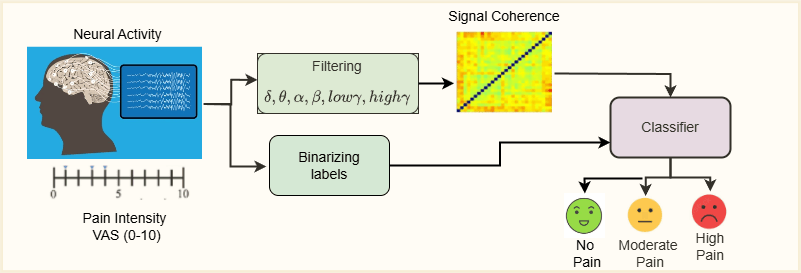}
        \label{fig:signal_coherence}
    }
    
    \caption{The (a) figure illustrates the model pipeline utilizing PIB values for ternary classification, \\while the (b) figure depicts the model pipeline employing Magnitude Squared Coherence (MSC) for the same classification task. For binary classification, we adopt a similar pipeline; however, we consolidate the moderate pain and high pain categories into a single "pain" class, with the "no pain" category designated as the "no pain" class.}
    \label{fig:pipeline}
\end{figure}

Each windowed signal is then bandpass filtered to isolate six specific frequency bands: \textit{delta} ($\delta$ (0.1 - 4 Hz)), \textit{theta} ($\theta$ (4 - 8 Hz)), \textit{alpha} ($\alpha$ (8 - 13 Hz), \textit{beta} ($\beta$ (13 - 30 Hz), \textit{gamma} ($\gamma$ (30 - 60 Hz), and \textit{high gamma} ($\gamma_{\text{high}}$ (60 - 200 Hz)). Feature extraction is performed on each band, where PIB values and MSC are calculated as pain-related features. These extracted features, along with their corresponding labels, are subsequently fed into a classifier for supervised learning. The overall pipeline is depicted in Figure \ref{fig:pipeline}.
\section{Experiments}
\label{sec:3}
\begin{figure}[ht]
    \centering
    \begin{tabular}{ccc}
    
    \subfigure[Subject 1]{
        \includegraphics[width=0.3\textwidth]{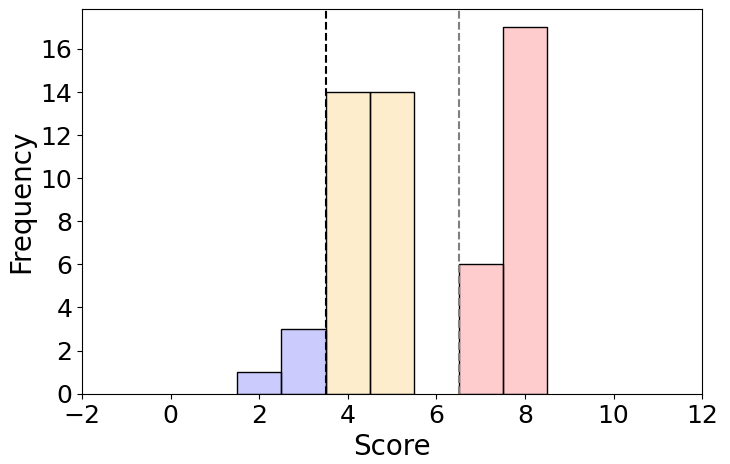}
        \label{fig:422bc5_label}
    }&
    \subfigure[Subject 2]{
        \includegraphics[width=0.3\textwidth]{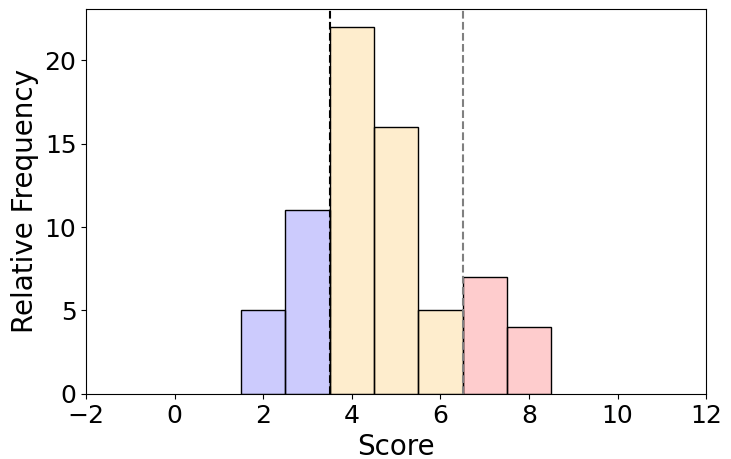}
        \label{fig:0b5a2e_label}
    }
    &
    \subfigure[Subject 3]{
        \includegraphics[width=0.3\textwidth]{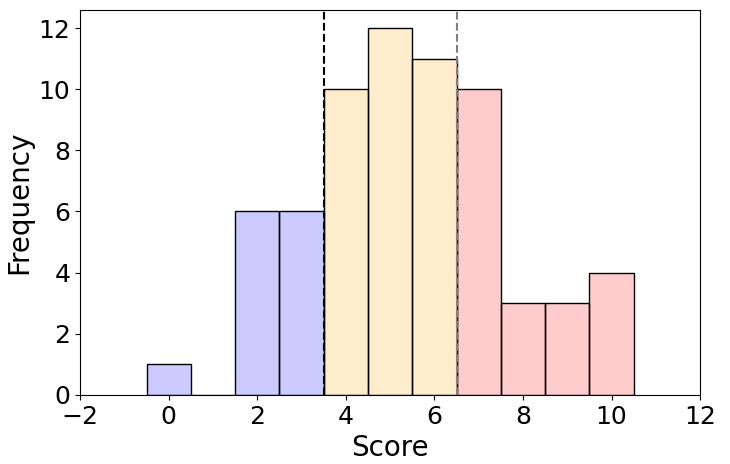}
        \label{fig:c5a5e9_label}
    }
    
    % Add the separate legend image below the main figure
    \end{tabular}
    \vspace{\baselineskip}
    \centering
    \includegraphics[width=1\textwidth]{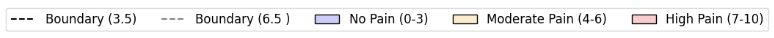} % Adjust width accordingly

    \caption{ The histogram of pain labels across various subjects is illustrated. For the binary classification task, "moderate pain" and "high pain" are combined into a single "pain" class.}
    \label{fig:pain_label_distribution}
\end{figure}

The pain data used in this study presents unique challenges due to its subjective nature and reliance on self-reported scores, making it inherently difficult to work with. Each individual’s pain experience varies significantly, and this variability adds to the complexity of accurately classifying pain levels.
The distribution of pain labels for each subject is shown in Figure \ref{fig:pain_label_distribution}.

% \begin{figure}[ht]
%     \centering
%     \begin{minipage}[b]{0.45\columnwidth} % Adjust to fit within a single column
%         \centering
%         \includegraphics[width=\textwidth]{422bc5_hist_label.png}
%         (Subject 1)
%         \label{fig:422bc5_label}
%     \end{minipage}
%     \hfill
%     \vskip\baselineskip % Space between rows

%     \begin{minipage}[b]{0.45\columnwidth}
%         \centering
%         \includegraphics[width=\textwidth]{0b5a2e_hist_label.png}
%         (Subject 2)
%         \label{fig:0b5a2e_label}
%     \end{minipage}
%     \hfill
%     \begin{minipage}[b]{0.45\columnwidth}
%         \centering
%         \includegraphics[width=\textwidth]{c5a5e9_hist_label.png}
%         (Subject 3)
%         \label{fig:c5a5e9_label}
%     \end{minipage}

%     % Add the separate legend image below the main figure
%     \vskip\baselineskip % Adjust space if necessary
%     \centering
%     \includegraphics[width = 1\textwidth]{legend.png} % Adjust width accordingly
%     \caption{The histogram of pain labels across various subjects is illustrated. For the binary classification task, "moderate pain" and "high pain" are combined into a single "pain" class.}
%     \label{fig:pain_label_distribution}
% \end{figure}

% Always give a unique label
% and use \ref{<label>} for cross-references
% and \cite{<label>} for bibliographic references
% use \sectionmark{}
% to alter or adjust the section heading in the running head
% Instead of simply listing headings of different levels we recommend to let every heading be followed by at least a short passage of text.  Further on please use the \LaTeX\ automatism for all your cross-references and citations as has already been described in Sect.~\ref{sec:2}.

% Please note that the first line of text that follows a heading is not indented, whereas the first lines of all subsequent paragraphs are.

% If you want to list definitions or the like we recommend to use the enhanced \verb|description| environment -- it will automatically rendered in line with the preferred layout.

% \begin{description}[Type 1]
% \item[Type 1]{That addresses central themes pertainng to migration, health, and disease. In Sect.~\ref{sec:1}, Wilson discusses the role of human migration in infectious disease distributions and patterns.}
% \item[Type 2]{That addresses central themes pertainng to migration, health, and disease. In Sect.~\ref{subsec:2}, Wilson discusses the role of human migration in infectious disease distributions and patterns.}
% \end{description}
In the binary and ternary classification pipeline, we first randomly selected 10\% of each subject's trials as test data to mitigate the risk of data leakage between the training and test sets. The remaining trials were then subjected to a 20-fold cross-validation strategy. In each fold, we trained a model on a randomly selected, class-balanced subset of trials, with each trial divided into 30 non-overlapping 10-second windows, followed by the feature extraction process. We experimented with three preprocessing methods: PIB, MSC, and the PIB + MSC combination. The trained model was then evaluated on the designated test data. This experimental procedure was repeated 15 times, yielding an accuracy distribution across the iterations. We report the mean accuracy for each subject.

% For both binary and ternary classification tasks, we utilize Scikit-learn's Logistic Regression \cite{logistic_sklearn} with an \( \ell_2 \) penalty to regularize model coefficients \( \beta \), combined with cross-entropy loss as our baseline classifier. Formally, the logistic loss for \( n \) samples is minimized as follows:
For both binary and ternary classification tasks, we utilize Scikit-learn's Logistic Regression \cite{logistic_sklearn} with an $l_{2}$ penalty to regularize model's coefficients $\beta$, combined with cross-entropy loss as our baseline classifier. Formally, the logistic loss for $\textit{n}$ samples is minimized as follows:

\begin{equation}
\mathcal{L}(\beta) = -\frac{1}{n} \sum_{i=1}^{n} \left[ y_i \log(p_i) + (1 - y_i) \log(1 - p_i) \right] + \lambda \|\beta\|_2^2,
\end{equation}

where \( p_i \) is the probability estimate for the \( i \)-th sample, and \( y_i \) represents the binary/ternary label.

We further evaluate the performance of Scikit-learn's Support Vector Machines (SVM) \cite{svc_sklearn} with an \( \ell_2 \) penalty and a radial basis function (RBF) kernel 
\begin{equation}
 K(x_i, x_j) = \exp\left(-\gamma \|x_i - x_j\|^2\right) 
\end{equation}
where \( \|x_i - x_j\| \) denotes the Euclidean distance between \( x_i \) and \( x_j \), and \( \gamma \) is a scaling parameter that controls the influence of individual data points on the decision boundary. A higher \( \gamma \) value causes the kernel to focus more narrowly on points close to \( x_i \), while a lower \( \gamma \) value allows for a broader influence, capturing more global relationships between data points.

Additionally, we employ Scikit-learn's Random Forest classifier (RF) \cite{rf_sklearn} with \( n_{\text{estimators}} = 100 \) decision trees, using the Gini criterion to measure impurity. Each tree selects a random subset of features at each split, where the number of features is determined as $\sqrt{d}$, with $\textit{d}$  being the total number of features. This square-root selection strategy increases model diversity by reducing feature correlation across trees, enhancing generalization.

\section{Results}
The results for both binary and ternary classifications for all the three strategies are presented in Tables \ref{tab:hyp1-hyp2-pib}-\ref{tab:ternary-combined}, respectively. The tables report the mean accuracy derived from the distribution of accuracy across the experimental trials.

We compared the performance of the models against the chance level accuracy. From the Tables \ref{tab:hyp1-hyp2-pib}-\ref{tab:ternary-combined}, it can be noted that most of the subjects performed comparable to or marginally better than chance level accuracy (50\% in binary classification and 33\% in ternary classification). Notably, subject 3 performed significantly above chance level accuracy. This might be attributed to better feature-label representation to the models.

\begin{table*}[htbp]
\centering
\begin{tabular}{|| c || c || c || c || c || c || c || c || c || c ||} 
 \hline
 \textbf{Subject} & \multicolumn{3}{c||}{\textbf{Strategy 1}} & \multicolumn{3}{c||}{\textbf{Strategy 2}} & \multicolumn{3}{c||}{\textbf{Strategy 3}} \\
 \hline
 & \textbf{LR} & \textbf{SVM} & \textbf{RF} & \textbf{LR} & \textbf{SVM} & \textbf{RF} & \textbf{LR} & \textbf{SVM} & \textbf{RF} \\
 \hline\hline
Subject 1 & 50 & 52 & 47  & 63 & \textbf{68} & 62 & 52 & 55 & 57 \\
 \hline
Subject 2 & 52 & 47 & 54  & 58 & 58 & 56 & 58 & 57 & \textbf{60} \\ 
 \hline
Subject 3 & 63 & 64 & \textbf{73}  & 63 & 60 & 55 & 69 & 69 & \textbf{73} \\
 \hline
\end{tabular}
\caption{Accuracy Comparison of SVM and RF in Binary Classification Using PIB Features for strategies 1, 2, and 3.}
\label{tab:hyp1-hyp2-pib}
\end{table*}

\begin{table*}[htbp]
\centering
\begin{tabular}{|| c || c || c || c || c || c || c || c || c || c ||} 
 \hline
 \textbf{Subject} & \multicolumn{3}{c||}{\textbf{Strategy 1}} & \multicolumn{3}{c||}{\textbf{Strategy 2}} & \multicolumn{3}{c||}{\textbf{Strategy 3}} \\
 \hline
 & \textbf{LR} & \textbf{SVM} & \textbf{RF} & \textbf{LR} & \textbf{SVM} & \textbf{RF} & \textbf{LR} & \textbf{SVM} & \textbf{RF} \\
 \hline\hline
Subject 1 & 40 & 46 & 49  & 51 & \textbf{55} & \textbf{55} & 46 & 47 & 43 \\
 \hline
Subject 2 & 47 & 46 & 46  & 55 & 57 & 56 & 58 & \textbf{64} & 60 \\
 \hline
Subject 3 & \textbf{76} & 75 & 65  & 61 & 61 & 62 & 67 & 72 & 69 \\
 \hline
\end{tabular}
\caption{Accuracy comparison of SVM and RF in binary classification using MSC features for strategies 1, 2, and 3.}
\label{tab:hyp1-hyp2-msc}
\end{table*}

\begin{table*}[htbp]
\centering
\begin{tabular}{|| c || c || c || c || c || c || c || c || c || c ||} 
 \hline
 \textbf{Subject} & \multicolumn{3}{c||}{\textbf{Strategy 1}} & \multicolumn{3}{c||}{\textbf{Strategy 2}} & \multicolumn{3}{c||}{\textbf{Strategy 3}} \\
 \hline
 & \textbf{LR} & \textbf{SVM} & \textbf{RF} & \textbf{LR} & \textbf{SVM} & \textbf{RF} & \textbf{LR} & \textbf{SVM} & \textbf{RF} \\
 \hline\hline
Subject 1 & 49 & 56 & 44  & \textbf{58} & 51 & 58 & 50 & 56 & 52 \\
 \hline
Subject 2 & 46 & 48 & 46  & 53 & 54 & \textbf{64} & 54 & 54 & 54 \\
 \hline
Subject 3 & 60 & 50 & \textbf{66}  & 50 & 49 & 52 & 63 & 50 & 71 \\
 \hline
\end{tabular}
\caption{Accuracy comparison of SVM and RF in binary classification using PIB + MSC features for strategies 1, 2, and 3. }
\label{tab:hyp1-hyp2-pib_msc}
\end{table*}

\begin{figure}[ht]
    \centering
    \includegraphics[width=0.6\linewidth]{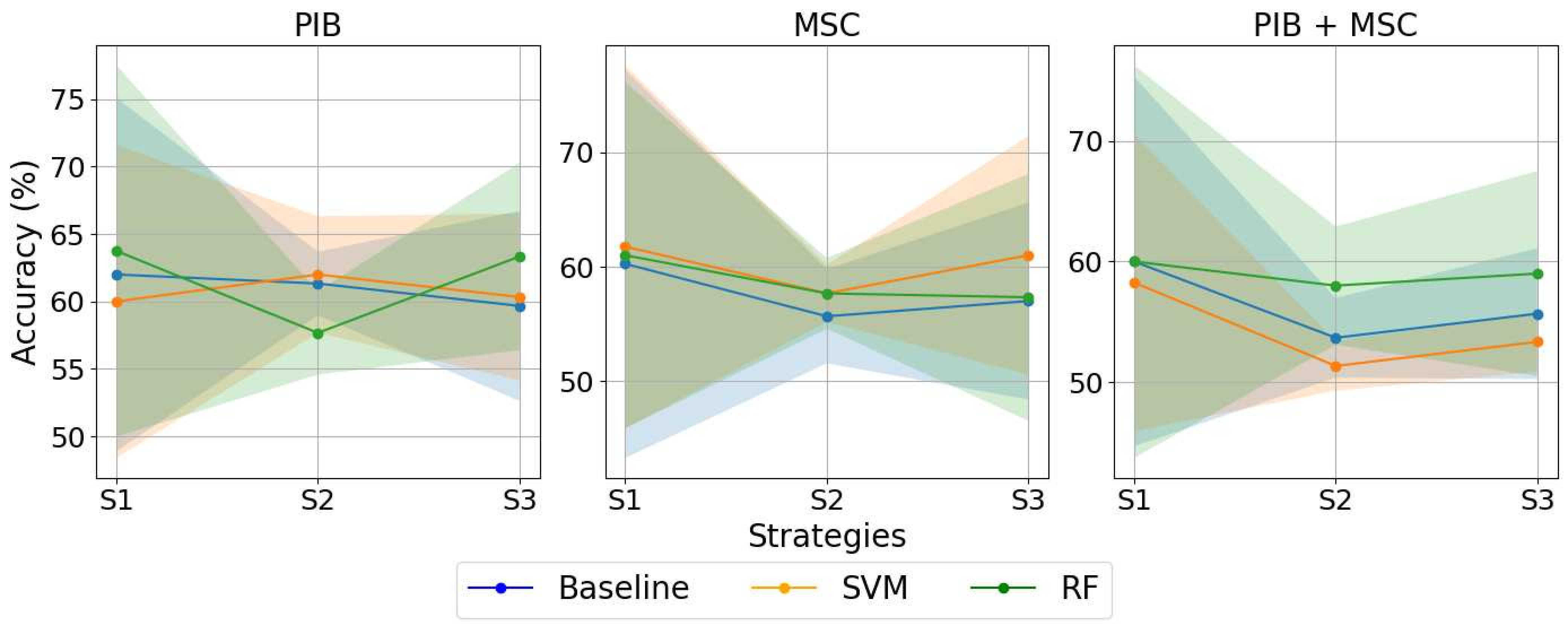}
    \caption{Accuracy trends across strategies for different models and features (with standard deviation). The plot illustrates the performance of various feature sets (PIB, MSC, PIB + MSC, Ternary PIB, Ternary MSC, Ternary PIB + MSC) across three different models (LR, SVM, RF) for three strategies 1, 2 and 3 denoted by S1, S2 and S3 respectively. Each subplot shows the mean accuracy for each model across all the subjects, with shaded regions representing the standard deviation. The trends demonstrate the models' stability and variance in performance across the strategies and feature sets.}
    \label{fig:acc_stats}
\end{figure}

\begin{table*}[htbp]
\centering
 % Resize to fit within text width
\begin{tabular}{||c||c||c||c||c||c||c||c||c||c||} 
 \hline
 \textbf{Subject} & \multicolumn{3}{c|}{\textbf{PIB}} & \multicolumn{3}{c|}{\textbf{MSC}} & \multicolumn{3}{c|}{\textbf{PIB + MSC}} \\
 \hline
 & \textbf{LR} & \textbf{SVM } & \textbf{RF } & \textbf{LR} & \textbf{SVM} & \textbf{RF} & \textbf{LR } & \textbf{SVM } & \textbf{RF } \\
 \hline
Subject 1 & 31 & 34 & 31 & 32 & 30 & 31 & 32 & \textbf{38} & 31 \\
 \hline
Subject 2 & 38 & 35 & 36 & 38 & \textbf{39} & 37 & 36 & 36 & 33 \\
 \hline
Subject 3 & 52 & 47 & 46 & \textbf{56} & 51 & 52 & 42 & 34 & 52 \\
 \hline
\end{tabular}

\caption{Accuracy Comparison of SVM and RF in \textbf{Ternary Classification} Using PIB, MSC, and PIB + MSC Features for Strategy 1.}
\label{tab:ternary-combined}
\end{table*}

Figure \ref{fig:acc_stats}  provides a visual summary of the statistics presented in Tables \ref{tab:hyp1-hyp2-pib}-\ref{tab:hyp1-hyp2-pib_msc}. Each subplot displays the mean accuracy across subjects for various strategies and feature sets. Upon inspection, it is evident that the PIB feature set consistently yields the best performance in binary classification tasks (also compare Tables \ref{tab:hyp1-hyp2-pib}, \ref{tab:hyp1-hyp2-msc}, \ref{tab:hyp1-hyp2-pib_msc}). Additionally, the RF model outperforms the other models (LR and SVM) in terms of mean accuracy, demonstrating its superior effectiveness for these tasks.

For ternary classification tasks (refer Table \ref{tab:ternary-combined}), it can be noted that MSC features provide better results compared to PIB and PIB + MSC, and LR model outperforms the other two models in ternary classification.

\begin{figure*}[htbp] % Use figure* for full-width figure in two-column layout
    % \centering
    \subfigure[(Pain network of Subject 1)]{
        \includegraphics[width=0.5\textwidth]{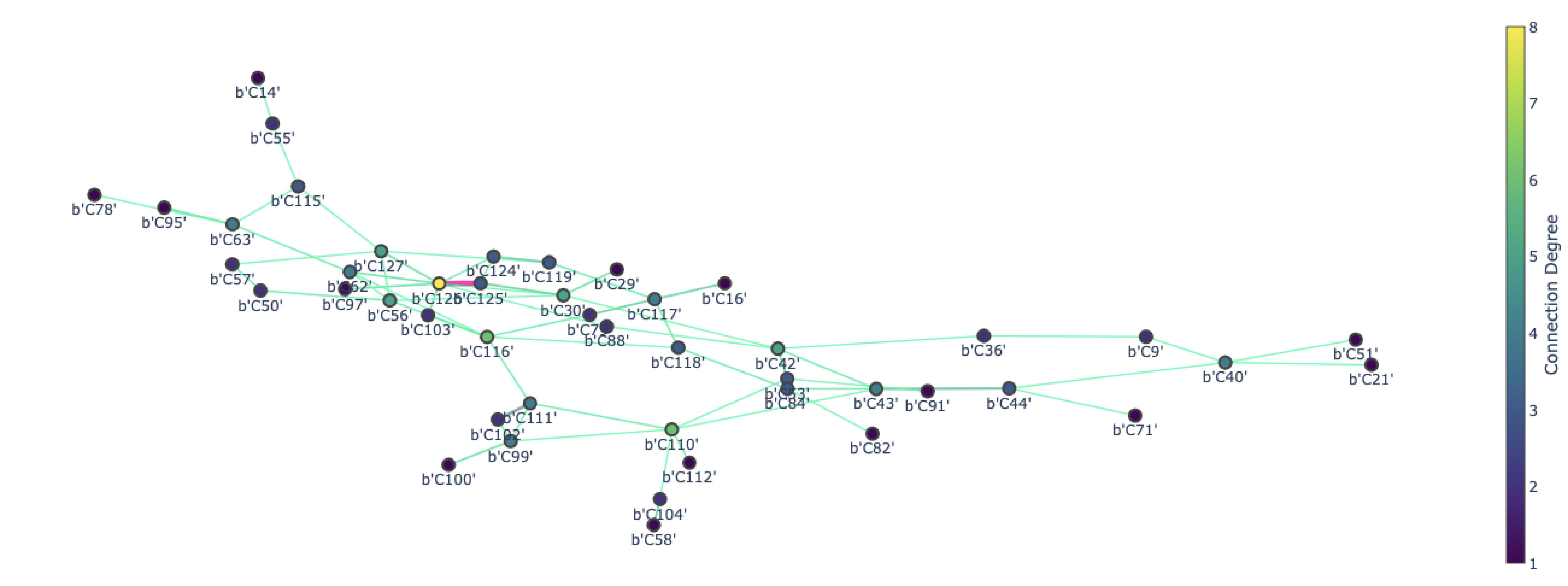}
        \label{fig:422bc5_pnw}
    }
    \hfill
    \subfigure[(Pain network of Subject 2)]{
        \includegraphics[width=0.5\textwidth]{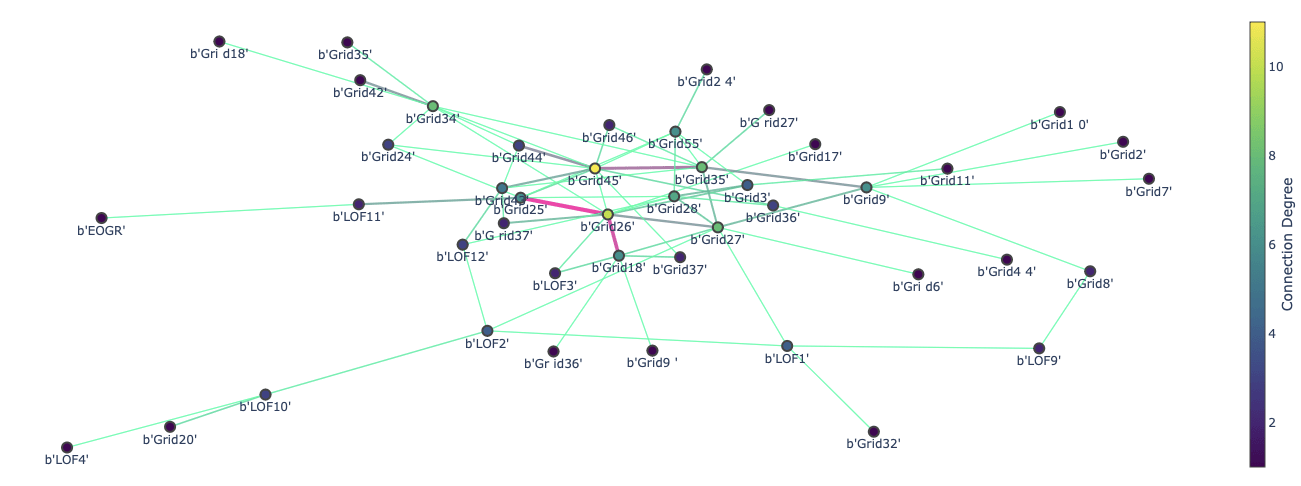}
        \label{fig:0b5a2e_pnw}
    }
    \hfill
    \subfigure[(Pain network of Subject 3)]{
        \includegraphics[width=0.5\textwidth]{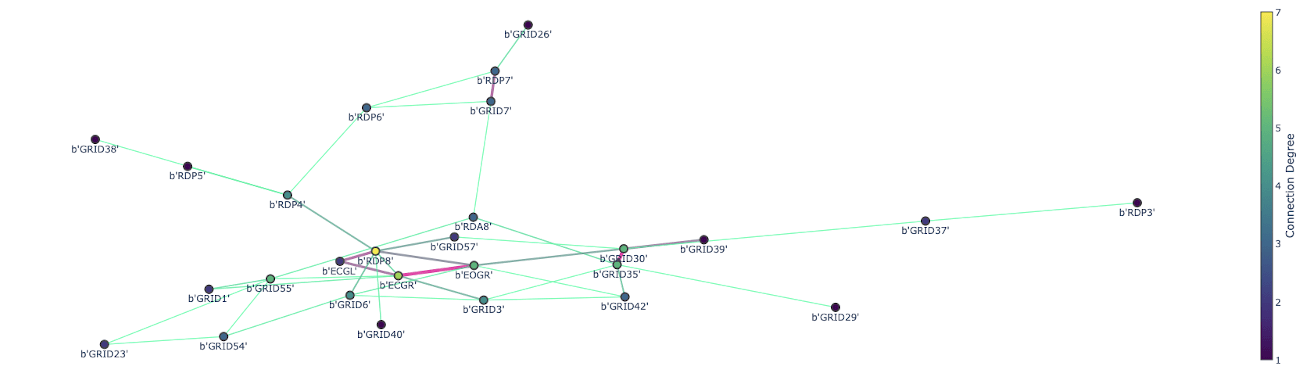}
        \label{fig:c5a5e9_pnw}
    }
    
    \caption{The pain network visualization illustrates the connections between electrodes, with edge thickness and color intensity reflecting the summed feature importance values from a random forest model in a binary pain classification task. The edges are colored such that more pinkish edges indicate a higher importance of the connection between the two electrodes. Node color represents the degree of each electrode, with more yellowish nodes signifying higher centrality within the network, as shown by the color bar. The top 3-4 electrodes with the highest degree are highlighted, emphasizing their significance in the classification task. This visualization highlights the most influential electrodes and their relationships within the pain network.}
    \label{fig:pain_networks_combined}
\end{figure*}

From the binary classification using MSC features under Strategy 1, we evaluated the most important electrode-electrode combinations based on Gini Impurity from the RF model. Subsequently, we constructed a "pain network" utilizing these significant features. The network is illustrated in Figure \ref{fig:pain_networks_combined}.

The Figure \ref{fig:pain_networks_combined} depicts the "pain network", a network of electrode connections where the edges are weighted by the summed feature importance values from a random forest classifier, trained to distinguish between pain and non-pain states. 
% Each edge in the network represents a connection between two electrodes, with the weight of the edge proportional to the importance of the connection for classification. The color of the edges is normalized based on their weight, with more intense colors indicating stronger connections, and the width of the edges further reflects this importance. The node color represents the degree of each electrode, i.e., the number of connections it has, helping to highlight the most central electrodes in the network.
The top 3-4 electrodes, which have the highest degree centrality, are emphasized to provide further insight into the key players in the pain network. This visualization provides a powerful representation of how electrode pairs and their relationships contribute to pain state detection.

\begin{figure}[ht]
    \centering
    \begin{tabular}{ccc}

    \subfigure[(Subject 1)]{
        \includegraphics[width=0.15\textwidth]{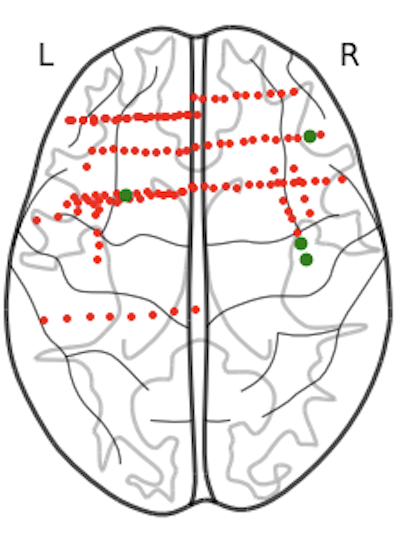}
        \label{fig:422bc5_brain_green}
    }
    &
    \subfigure[(Subject 2)]{
        \includegraphics[width=0.15\textwidth]{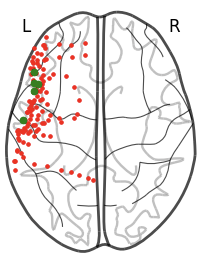}
        \label{fig:0b5a2e_brain_green}
    }
    &
    \subfigure[(Subject 3)]{
        \includegraphics[width=0.15\textwidth]{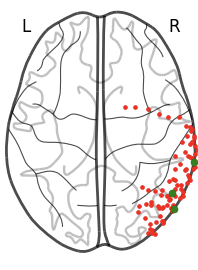}
        \label{fig:c5a5e9_brain_green}
    }
    \end{tabular}
    \caption{Axial view of electrode distribution across subjects, with key channels contributing to pain classification highlighted in green.}
    \label{fig:electrode_location_green}
\end{figure}

%%%%%%%%%%%%%%%%%%%%%%%%%%%%%%%%%%%
For Subject 1, key channels for distinguishing "pain" from "no pain" include coherence signals between the posterior division of the Middle Temporal Gyrus, the anterior division of the Parahippocampal Gyrus, the Middle Frontal Gyrus, and the Superior Frontal Gyrus, respectively. 
In Subject 2, the Middle Frontal Gyrus, Precentral Gyrus, and Superior frontal—were identified as critical. For Subject 3, the primary areas identified include the temporo-occipital part of the Middle Temporal Gyrus, the posterior division of the Supramarginal Gyrus, and the superior division of the Lateral Occipital Cortex. The majority of the regions mentioned above align with those identified in the work \cite{Rockholt2023Jun}. These regions, highlighted in the brain montage plots shown in Figure \ref{fig:electrode_location_green}, represent significant areas contributing to pain classification.

\section{Conclusion and Future Work}
In this work, we propose a systematic machine learning-based approach for binary and ternary classification of acute pain from raw electrophysiological signals, utilizing PIB and signal coherence, a functional connectivity metric, as key features. We explored the use of traditional machine learning algorithms on a manually curated electrophysiological dataset and compared their performance against chance levels. Subjects 1 and 2 demonstrated marginally better-than-chance performance in both binary and ternary classification, whereas Subject 3 achieved substantially better results. The results exhibited significant variation across patients, potentially due to the efficacy of feature-label representation. Furthermore, the pain scores were self-reported by patients, introducing subjectivity that may have contributed to the high variance observed.

Despite the patients having electrodes at varying locations and reporting diverse pain scores, we successfully developed a common methodology that achieved accuracy consistently better than chance. Additionally, we identified critical electrode pairings associated with acute pain detection, shedding light on the specific neural markers that distinguish pain states.

% We hope this methodology will inspire further research and adoption within the neuroscience and machine learning communities to advance research in understanding and classifying pain states.

% \section{Future Work}
Future work includes expanding the feature set to include diverse neural and physiological metrics, collecting larger and more diverse datasets for improved generalizability, and refining classification to capture finer-grained pain levels, enabling more precise and personalized pain assessment.
% \section{Acknowledgments}
% This work was supported by the Department of Electrical and Computer Engineering, Paul G. Allen School of Computer Science and Engineering,  members of the Neural Systems Lab, and Herron Lab at the University of Washington, Seattle. We would also like to thank Harborview Medical Center for allowing us to collect the data used in this research.
\section{Acknowledgement and Ethics Approval}
This work was supported by the Department of Electrical and Computer Engineering, Paul G. Allen School of Computer Science and Engineering, members of the Neural Systems Lab, and Herron Lab at the University of Washington, Seattle. We would also like to thank Harborview Medical Center for allowing us to collect the data used in this research. This work was funded in part by the National Science Foundation (NSF) EFRI grant no. 2223495, a Weill Neurohub Investigator grant, and a CJ and Elizabeth Hwang endowed professorship (RPNR).

% \ethics{Competing Interests}{Please declare any competing interests
% in the context of your chapter. The following sentences can be regarded as examples.\newline
% This study was funded by [X] [grant number X].\newline
%  [Author A] has a received research grant from [Company W].\newline
%  [Author B] has received a speaker honorarium from [Company X] and owns stock in [Company~Y].\newline 
%  [Author C] is a member of [committee Z].\newline 
% The authors have no conflicts of interest to declare that are relevant to the content of this chapter.}

% \eject

All experiments involving human participants were approved by the University’s Institutional Review Board (IRB), and informed consent was obtained from all participants.
\bibliographystyle{unsrtnat}
\bibliography{references.bib}

\section*{Appendix} 
\addcontentsline{toc}{section}{Appendix}
\section{Kernel Density Estimation (KDE) of the pain score distribution}
\begin{figure}[ht]
    \centering
    \begin{tabular}{ccc}
    \subfigure[Subject 1]{
        \includegraphics[width=0.25\columnwidth]{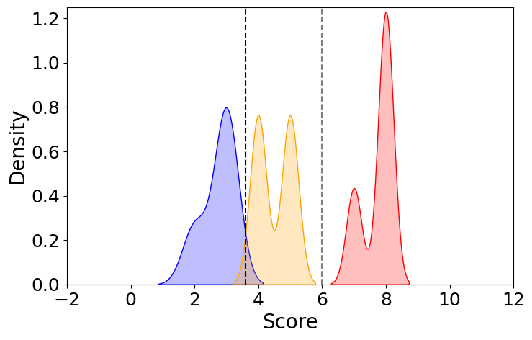}
        \label{fig:422bc5_kde}
    }&
    % \hfill
    \subfigure[Subject 2]{
        \includegraphics[width=0.25\columnwidth]{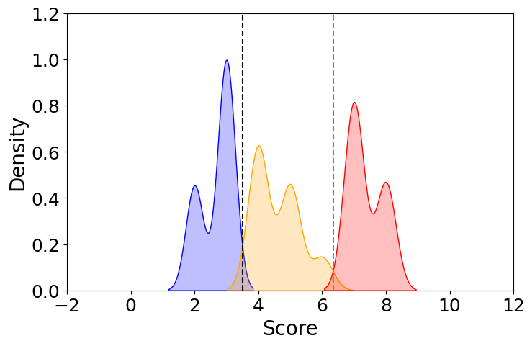}
        \label{fig:0b5a2e_kde}
    }&
    % \vskip\baselineskip % Adjust space if necessary
    \subfigure[Subject 3]{
        \includegraphics[width=0.25\columnwidth]{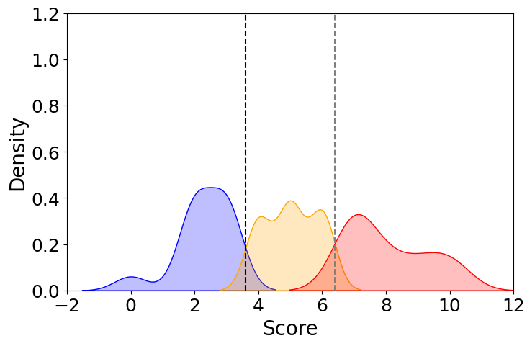}
        \label{fig:c5a5e9_kde}
    }
    \end{tabular}
    % Add the separate legend image below the main figure
    \vskip\baselineskip % Adjust space if necessary
    \centering
    \includegraphics[width=1\textwidth]{legend.png} % Adjust width accordingly
    \caption{The distribution of pain labels across various subjects is illustrated. For the binary classification task, "moderate pain" and "high pain" are combined into a single "pain" class.}
    \label{fig:pain_label_kde}
\end{figure}

The kernel density estimation of the pain scores for various subjects is provided in Figure \ref{fig:pain_label_kde}
\subsection{Top 10 electrodes for binary pain classification}
The top 10 electrodes, determined using the method described in the Results section, are shown in Table \ref{top-10-electrodes}.
\begin{table}[htbp]
\centering
\begin{tabular}{|c|c|c|}
\hline
\textbf{Subject} & \textbf{Electrode (Channel)} & \textbf{Brain Region} \\ \hline
\multirow{10}{*}{Subject 1} & RAMTG9 & Middle temporal gyrus posterior division \\ \cline{2-3} 
& RAMTG3 & Parahippocampal gyrus anterior division \\ \cline{2-3} 
& RDACC11 & Middle frontal gyrus \\ \cline{2-3} 
& LDACC7 & Superior frontal gyrus \\ \cline{2-3} 
& LPMFG7 & Superior frontal gyrus \\ \cline{2-3} 
& LAMTG13 & Temporal pole \\ \cline{2-3} 
& ROPER1 & Insular cortex \\ \cline{2-3} 
& LDACC9 & Middle frontal gyrus \\ \cline{2-3} 
& RAMTG5 & Parahippocampal gyrus anterior division \\ \cline{2-3} 
& LSI1 & Precentral gyrus \\ \hline
% \multirow{10}{*}{Subject 2} & GRID57 & Middle frontal gyrus \\ \cline{2-3} 
% & GRID26 & Inferior frontal gyrus pars triangularis \\ \cline{2-3} 
% & GRID5 & Middle temporal gyrus posterior division \\ \cline{2-3} 
% & GRID28 & Precentral gyrus \\ \cline{2-3} 
% & GRID61 & Precentral gyrus \\ \cline{2-3} 
% & GRID27 & Inferior frontal gyrus pars triangularis \\ \cline{2-3} 
% & ROF3 & Frontal pole \\ \cline{2-3} 
% & GRID40 & Supramarginal gyrus posterior division \\ \cline{2-3} 
% & GRID60 & Middle frontal gyrus \\ \cline{2-3} 
% & GRID18 & Temporal pole \\ \hline
\multirow{10}{*}{Subject 3} & GRID45 & Middle Frontal Gyrus \\ \cline{2-3} 
& GRID26 & Precentral Gyrus \\ \cline{2-3} 
& GRID35 & Middle Frontal Gyrus \\ \cline{2-3} 
& GRID34 & Superior Frontal Gyrus \\ \cline{2-3} 
& GRID27 & Middle Frontal Gyrus \\ \cline{2-3} 
& GRID28 & Middle Frontal Gyrus \\ \cline{2-3} 
& GRID25 & Superior Frontal Gyrus \\ \cline{2-3} 
& GRID18 & Inferior Frontal Gyrus \\ \cline{2-3} 
& GRID9 & Middle Frontal Gyrus \\ \cline{2-3} 
& GRID55 & Middle Frontal Gyrus \\ \hline
\multirow{10}{*}{Subject 4} & RDP8 & Middle temporal gyrus temporo-ocipital part \\ \cline{2-3} 
& GRID35 & Supramarginal gyrus posterior division \\ \cline{2-3} 
& GRID30 & Lateral occipital cortex superior division \\ \cline{2-3} 
& GRID55 & Lateral occipital cortex superior division \\ \cline{2-3} 
& GRID3 & Inferior temporal gyrus temporo-ocipital part \\ \cline{2-3} 
& RDP4 & Lateral occipital cortex inferior division \\ \cline{2-3} 
& GRID6 & Lateral occipital cortex inferior division \\ \cline{2-3} 
& RDP7 & Lateral occipital cortex inferior division \\ \cline{2-3} 
& GRID7 & Lateral occipital cortex inferior division \\ \cline{2-3} 
& GRID42 & Supramarginal gyrus anterior division \\ \hline
\end{tabular}
\caption{Top 10 Electrodes and Corresponding Brain Regions for Each Subject}
\label{top-10-electrodes}
\end{table}

\section{Algorithm for pain classification}
The classification task is performed using a cross-validation approach to predict the pain levels based on the given data. The algorithm utilizes multiple iterations and folds to ensure robust evaluation. The steps involved in the classification process are described as follows:

\begin{enumerate}
    \item \textbf{Data Preparation:} The total dataset, consisting of N trials, is initially divided. A 10\% subset is randomly selected as the test set, while the remaining data is used for training.
    
    \item \textbf{Class Distribution:} The number of data points from each pain class (no pain, moderate pain, high pain) is determined. If the high pain class is available, its data points are also considered. The minimum number of data points across all classes is selected to ensure balanced representation in the training set.

    \item \textbf{Iteration and Cross-validation:} The algorithm performs a fixed number of iterations (\texttt{num\_iter}), where each iteration consists of \texttt{num\_folds} folds for cross-validation. In each fold, a random sample of data points from each class is selected, and the model is trained on these samples. The model's performance is evaluated on the test set, and the accuracy is recorded.

    \item \textbf{Model Training and Testing:} In each fold, the model is trained on the selected training data, and its accuracy is evaluated on the test data. The accuracy for each fold is stored for later analysis.

    \item \textbf{Final Accuracy Calculation:} After completing all folds for each iteration, the mean accuracy for that iteration is computed. Finally, the overall mean accuracy for the classification task is calculated by averaging the accuracies from all iterations.
\end{enumerate}
This algorithm provides a robust approach to evaluate the classification model's performance, ensuring generalizability across different subsets of the data. By using cross-validation and random sampling, the approach mitigates overfitting and assesses the model's ability to generalize to unseen data. The method is summarized in algorithm, which outlines the steps described above.

\newpage
\begin{tcolorbox}[title=Algorithm]
\begin{lstlisting}

# Algorithm: Cross-validation for Classification Task

# Input: Full dataset (trials, features)
# Output: Mean classification accuracy across all iterations

# Initialize:
test_set = []  # Test data selected for each iteration
num_iterations = 15  # Number of iterations for cross-validation
num_folds = 20  # Number of folds for each iteration
accuracy_per_iteration = []  # Stores accuracy for each iteration

# For iteration = 1 to num_iterations:
For iteration = 1 to num_iterations: 
    # Select a random 10% for testing
    test_set.append(random_subset(full_dataset, 10%))  
    # Remaining data after testing data selection
    training_set = remaining_dataset  

    # Get the number of samples for 'no pain' class
    num_no_pain_samples = get_num_samples(no_pain_class_data)  
    # Get the number of samples for 'moderate pain' class
    num_moderate_pain_samples = get_num_samples(moderate_pain_class_data)

    # Check if high_pain_class_data exists
    If high_pain_class_data exists: 
        # Get the number of samples for 'high pain' class
        num_high_pain_samples = get_num_samples(high_pain_class_data)  
    EndIf

    # Limit by the smallest class size
    max_samples_per_class = min(num_no_pain_samples,
                                num_moderate_pain_samples, 
                                num_high_pain_samples) 
                                
    # Stores accuracy for each fold in the current iteration
    accuracy_per_fold = []  
    
    # For fold = 1 to num_folds:
    For fold = 1 to num_folds:
        # Randomly select equal samples for each class
        no_pain_fold, moderate_pain_fold, high_pain_fold = get_random_subset(
        size = max_samples_per_class)  
        # Combine the folds for training
        training_data = [no_pain_fold, moderate_pain_fold, high_pain_fold]  
        # Train the model with the current fold's data
        train_model(training_data)

        # Test the model on the test set and store the accuracy
        accuracy_per_fold.append(test_model(test_set))  
    EndFor

    # Store mean accuracy for the current iteration
    accuracy_per_iteration.append(mean(accuracy_per_fold))  
EndFor

# Compute overall mean accuracy across all iterations
mean_accuracy = mean(accuracy_per_iteration)  
Return: mean_accuracy

\end{lstlisting}
\end{tcolorbox}

\section{Details of experimentation of Subject 4}
\subsection{Electrode coverage:}
\begin{figure}[ht]
    % \centering
    \centering
    \includegraphics[width=0.15\textwidth]{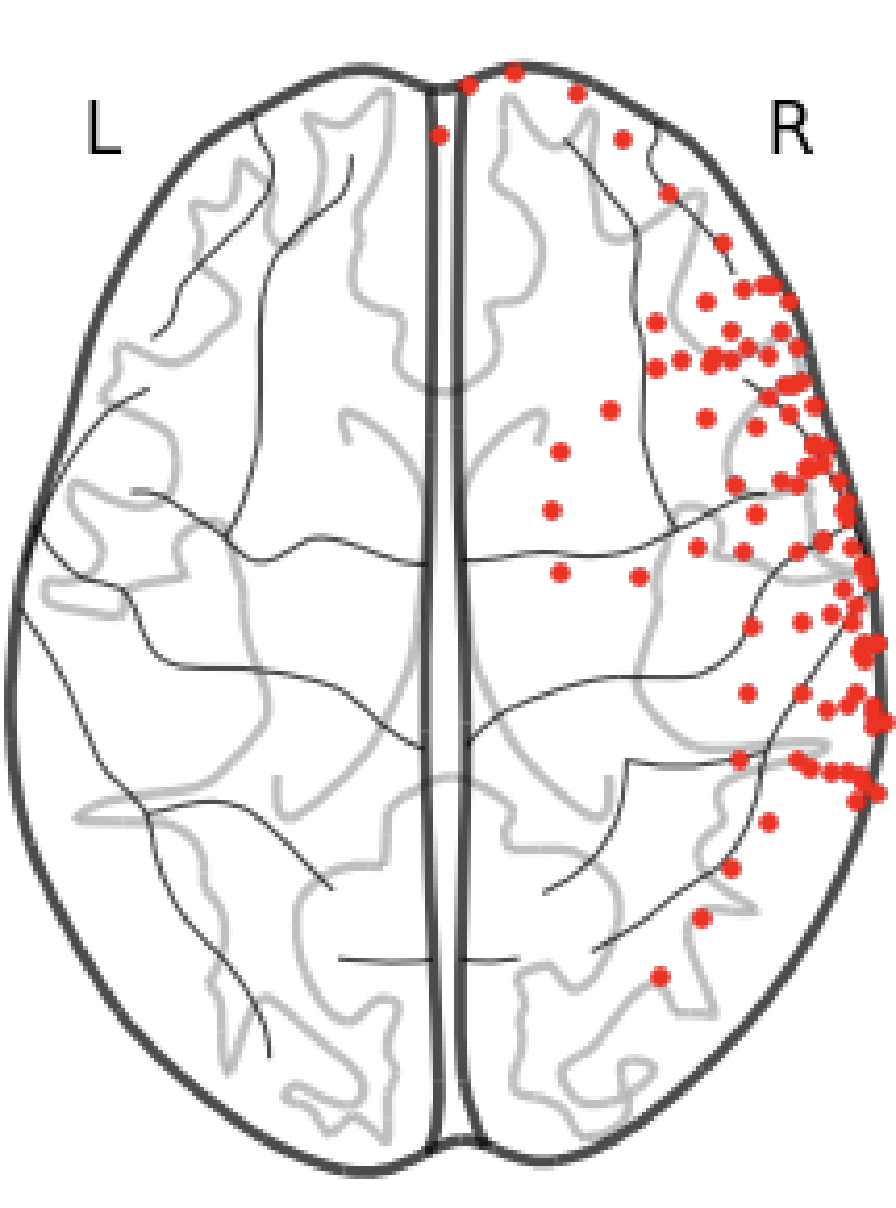}
    \caption{Axial view of electrodes for Subject 4}
    \label{fig:822e28_elec}
\end{figure}

The electrode coverage for Subject 4 is shown in Figure \ref{fig:822e28_elec}
The Subject 4 has surface ECoG electrodes concentrated in a single hemisphere like Subjects 2 and 3. 

\subsection{Dataset description for Subject 4:}

The description of data for Subject 4 is shown in Table \ref{tab:dataset_sub_4}.
\begin{table}[ht]
\centering
\begin{tabular}{|| c || c || c || c || c  || c ||} 
 \hline
 \textbf{ID} & \textbf{Total} & \textbf{Useful} & \textbf{Trials} & \textbf{Modality}  \\ 
 & \textbf{channels} & \textbf{channels} & &\\
 \hline\hline

  Subject 4 &129 &94 & 36& ECoG\\
  \hline

\end{tabular}
   \caption{Dataset description for Subject 4}
  \label{tab:dataset_sub_4}
\end{table} 

\subsection{Pain label distribution for Subject 4: }
\begin{figure}[htbp]
    \centering
    % Subfigure: Histogram of pain labels
    \subfigure[Histogram of pain labels]{
        \includegraphics[width=0.25\textwidth]{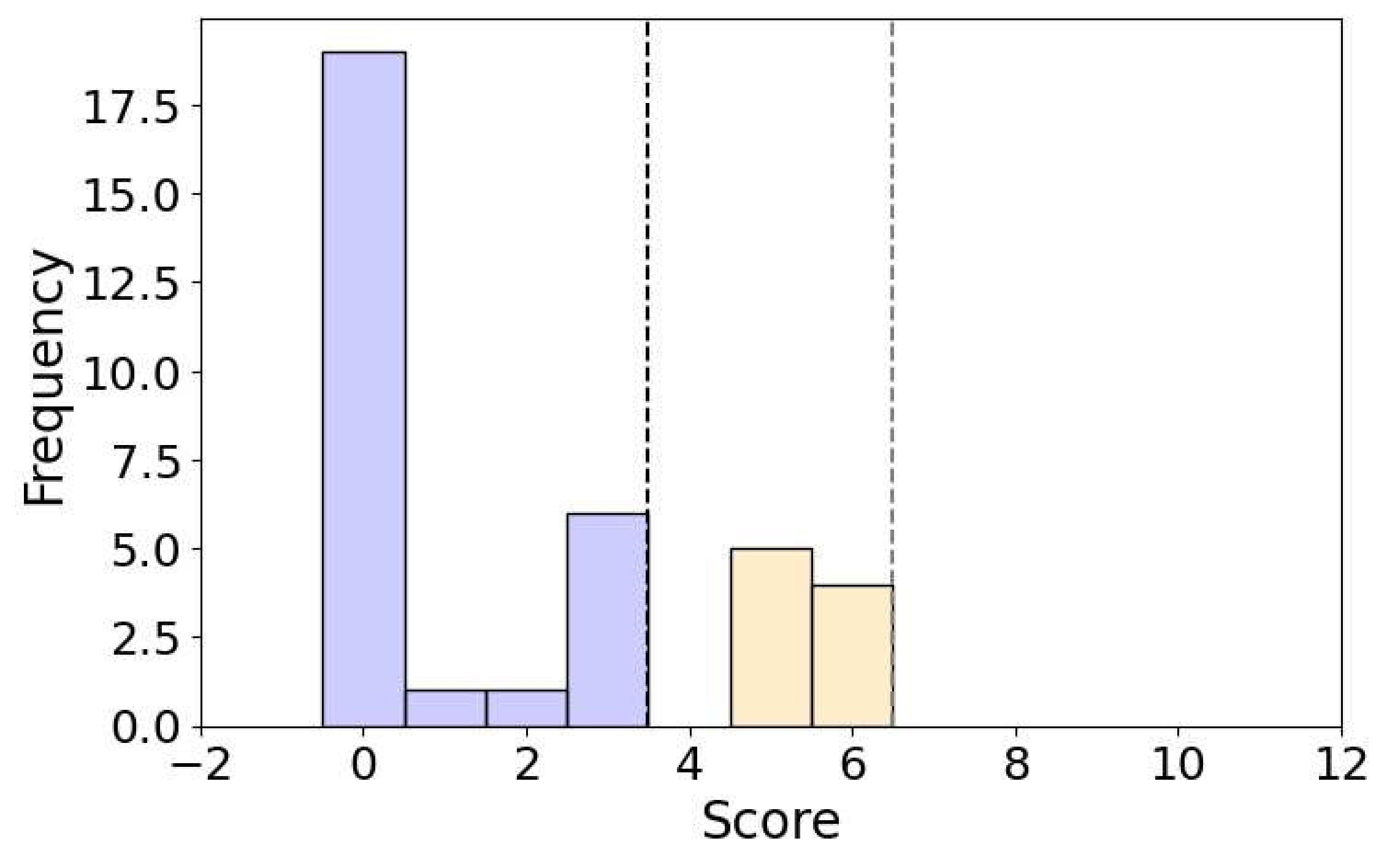}
        \label{fig:822e28_label}
    }
    % \hfill
    % Subfigure: KDE of pain labels
    \subfigure[KDE of pain labels]{
        \includegraphics[width=0.25\textwidth]{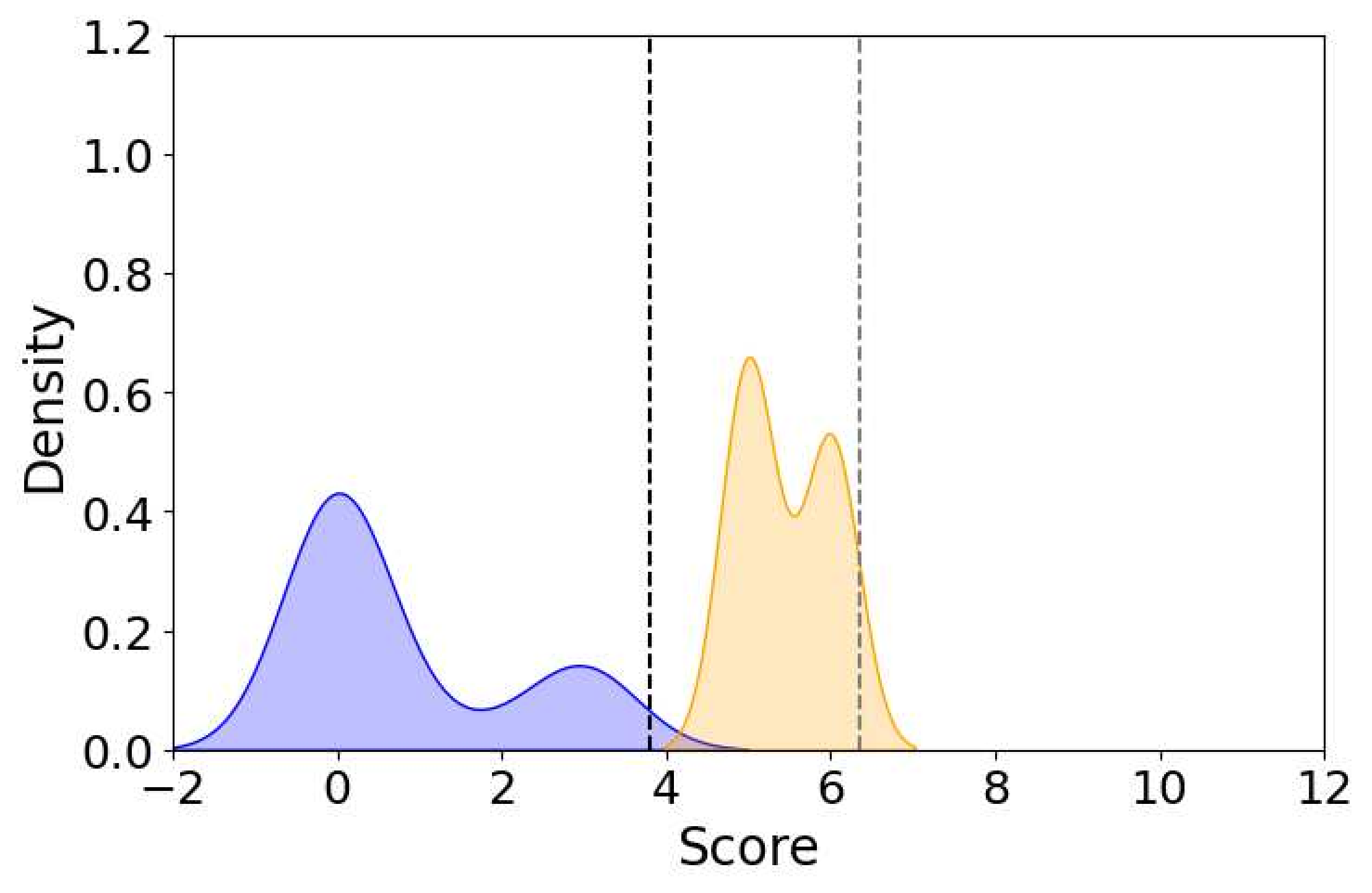}
        \label{fig:822e28_kde}
    }
    
    % Legend
    \vskip\baselineskip
    \includegraphics[width=1\textwidth]{legend.png}
    
    % Figure caption
    \caption{The histogram and KDE distribution of pain labels for Subject 4.}
    \label{fig:sub4_pain_label}
\end{figure}

The distribution of pain labels for Subject 4 is shown in Figure \ref{fig:sub4_pain_label}. Note that the subject did not report any high pain labels, and hence all the experiments for this subject were restricted to binary framework.

\subsection{Results for Subject 4: }
The experimental procedures for Subject 4 were identical to those conducted for the other subjects. However, since Subject 4 did not report any high pain labels, only Strategy 1 was applicable to their data.

\begin{table*}[htbp]
\centering
\begin{tabular}{|| c || c || c || c || c || c || c || c || c || c ||} 
 \hline
 \textbf{Features} & \textbf{LR} & \textbf{SVM} & \textbf{RF} \\
 \hline\hline

\multirow{1}{*}{PIB}  & 83 & 77 & 81 \\
 \hline
\multirow{1}{*}{MSC} & 78 & 80 & 84 \\
 \hline
\multirow{1}{*}{PIB + MSC} & 85 & 79 & 84 \\
 \hline

\end{tabular}
\caption{Accuracy comparison of SVM and RF in binary classification for Subject 4 using PIB, MSC, and PIB + MSC features. Since strategies 2 and 3 are not applicable, only results for Strategy 1 are shown.}
\label{tab:combined-results-sub4}
\end{table*}

\subsection{Identifying brain regions corresponding to the distinction between no pain and pain: }
The experimental protocols used to identify brain regions corresponding to the distinction between no pain and pain were identical to those followed for the other subjects.

For Subject 4, the most influential areas are the Middle Frontal Gyrus, the pars triangularis of the Middle Frontal Gyrus, the posterior division of the Middle Temporal Gyrus, and the Precentral Gyrus.

\begin{table}[htbp]
\centering
\begin{tabular}{|c|c|c|}
\hline
\textbf{Subject} & \textbf{Electrode (Channel)} & \textbf{Brain Region} \\ \hline

\multirow{10}{*}{Subject 4} & GRID57 & Middle frontal gyrus \\ \cline{2-3} 
& GRID26 & Inferior frontal gyrus pars triangularis \\ \cline{2-3} 
& GRID5 & Middle temporal gyrus posterior division \\ \cline{2-3} 
& GRID28 & Precentral gyrus \\ \cline{2-3} 
& GRID61 & Precentral gyrus \\ \cline{2-3} 
& GRID27 & Inferior frontal gyrus pars triangularis \\ \cline{2-3} 
& ROF3 & Frontal pole \\ \cline{2-3} 
& GRID40 & Supramarginal gyrus posterior division \\ \cline{2-3} 
& GRID60 & Middle frontal gyrus \\ \cline{2-3} 
& GRID18 & Temporal pole \\ \hline
\end{tabular}
\caption{Top 10 Electrodes and Corresponding Brain Regions for Subject 4}
\label{tab:top-10-electrodes-sub4}
\end{table}

The top 10 electrodes and corresponding brain regions for this subject are shown in table \ref{tab:top-10-electrodes-sub4}. Additionally, the top 4 electrodes are highlighted in green in the axial view of the electrode coverage plot in Figure \ref{fig:combined_pnw_brain} (b).

\subsubsection{Pain network visualization for Subject 4:}
\begin{figure}[ht]
    \centering
    \begin{tabular}{ccc}
    \subfigure[Pain network]{
        \includegraphics[width=0.45\textwidth]{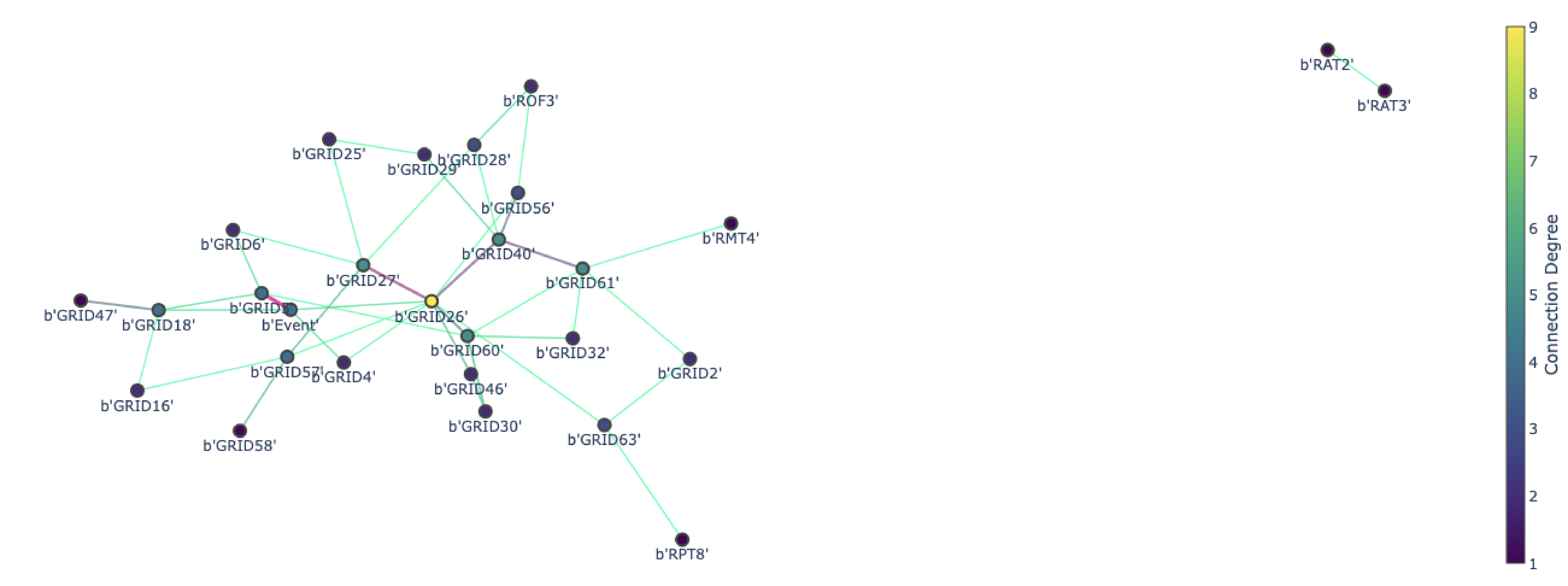}
        \label{fig:band_pnw}
    }
&
    \subfigure[Axial veiw of electrode coverage]{
        \includegraphics[width=0.15\textwidth]{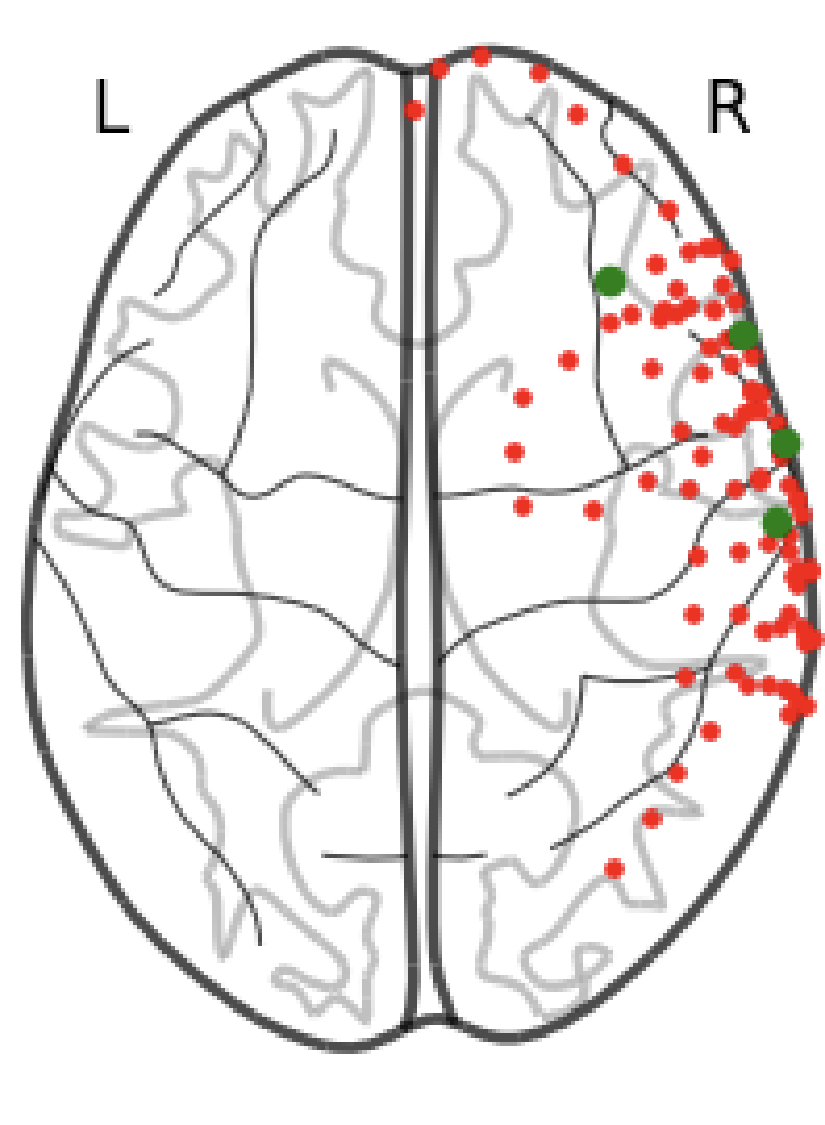}
        \label{fig:822e28_brain_green}
    }
    \end{tabular}
    \caption{(a) Pain network visualization and (b) axial view of electrode coverage for Subject 4.}
    \label{fig:combined_pnw_brain}
\end{figure}

The pain network for Subject 4 is shown in Figure \ref{fig:combined_pnw_brain} (a)

\end{document}